\documentclass[12pt]{article}
\usepackage{graphics}
\usepackage{epsfig,amsmath}
\input epsf

\begin{document}
\thispagestyle{empty}
\begin{center}

{\Large\bf{The extension to the transverse momentum
\vskip0.3cm
of the statistical parton distributions}}
\vskip1.4cm
{\bf Claude Bourrely and Jacques Soffer}  
\vskip 0.3cm
Centre de Physique Th\'eorique, UMR 6207 \footnote{ UMR 6207 is Unit\'e Mixte de Recherche du CNRS and of Universit\'es
Aix-Marseille I and Aix-Marseille II and of Universit\'e du Sud
Toulon-Var, laboratoire affili\'e \`a la FRUMAN.},\\
CNRS-Luminy, Case 907\\
F-13288 Marseille Cedex 9 - France \\ 
\vskip 0.5cm
{\bf Franco Buccella}
\vskip 0.3cm
Dipartimento di Scienze Fisiche, Universit\`a di Napoli,\\
Via Cintia, I-80126, Napoli
and INFN, Sezione di Napoli, Italy
\vskip 2cm
{\bf Abstract}\end{center}
By extending the statistical distributions to the transverse
degree of freedhom, we account for a multiplicative factor in the Fermi-Dirac functions of the light quarks, we 
were led to introduce in a previous work to comply with experiment.
We can also get light antiquark distributions, similar to those
we proposed earlier.\\

\vskip 1cm 
\noindent {\it keywords:}~polarized partons; transverse momentum;
statistical models

\vskip 1cm
\noindent CPT-2005/P.036\\
\noindent UNIV. NAPLES DSF-15/2005
\newpage
Resulting from a research which begun many years ago \cite{B2M2ST}, we proposed for 
the parton distributions \cite{BBS1}, some expressions inspired by quantum statistics, with a 
relationship between the light $q$ and $\bar{q}$ distributions following from
the chiral properties of QCD \cite{Bha}.
We determined the few free parameters involved, from a selected set of
precise deep inelastic scattering data \cite{BBS1}. Our approach has a strong predictive 
power, because, once the parameters, occuring in the better known 
unpolarized $u(x)$ and $d(x)$ distributions, are fixed, the polarized distributions, 
$\Delta u(x)$ and $\Delta d(x)$, as well as the antiquark distributions $\bar{u}(x)$, $\bar{d}(x)$, 
$\Delta \bar{u}(x)$ and $\Delta \bar{d}(x)$, are also fixed.\\
First, we made a prediction for a positive $A^n_1(x,Q^2)$ at high $x$ \cite{BBS2}, 
which was successfully verified by recent data at Jefferson Lab 
\cite{jlab02}. Next, we predicted for the high $x$ behavior of the ratio $\bar{d}(x)/\bar{u}(x)$
a monotonously increasing function of $x$. An earlier result from E866 at FNAL \cite{E866}, seemed
to contradict our prediction, but the comparison with the most recent 
results on Drell-Yan processes by the NUSEA \cite{E866a}, shows that this is not the case \cite{BBS3}.
Finally, the agreement with the data on $e^{\pm}p$ neutral and charge 
current reactions \cite{h100} \cite{zeus03}, is particularly successful. In Ref.~\cite{BBS3}
we stressed the experimental evidence in several structure functions for a 
typical property of the statistical distributions: the change of slope for $x>X^+_{0u}$, where
$X^+_{0u}$ denotes the largest ``thermodynamical potential'', whose definition and value
are given below.\\
We now recall some of the basic features of the statistical approach.
The fermion distributions are given by the sum of two terms \cite{BBS1},
a quasi Fermi-Dirac function and a helicity independent diffractive
contribution equal for all light quarks:
\begin{equation}
xq^h(x,Q^2_0)=
\frac{AX^h_{0q}x^b}{\exp [(x-X^h_{0q})/\bar{x}]+1}+
\frac{\tilde{A}x^{\tilde{b}}}{\exp(x/\bar{x})+1}~,
\label{equation1}
\end{equation}
\begin{equation}
x\bar{q}^h(x,Q^2_0)=
\frac{{\bar A}(X^{-h}_{0q})^{-1}x^{2b}}{\exp [(x+X^{-h}_{0q})/\bar{x}]+1}+
\frac{\tilde{A}x^{\tilde{b}}}{\exp(x/\bar{x})+1}~,
\label{equation2}
\end{equation}
at the input energy scale $Q_0^2=4 \mbox{GeV}^2$.\\
The parameter $\bar{x}$ plays the role of a {\it universal temperature}
and $X^{\pm}_{0q}$ are the two {\it thermodynamical potentials} of the quark $q$, with helicity $h=\pm$. 
For the gluons we consider the black-body inspired expression
\begin{equation}
xG(x,Q^2_0)=
\frac{A_Gx^{b_G}}{\exp(x/\bar{x})-1}~,
\label{equation3}
\end{equation}
a quasi Bose-Einstein function with $b_G=\tilde{b}+1$, since we believe that the diffractive
contribution in Eqs. (1,2) is strongly related to the gluons.\\
We also assume that, at the input energy scale, the polarized gluon distribution
vanishes, so
\begin{equation}
x\Delta G(x,Q^2_0)=0~.
\label{equation4}
\end{equation}
For strange quarks we assumed that $s=\bar{s}$, but we did not introduce any more 
parameters and we simply related them
to $\bar{u}+\bar{d}$ to comply with experimental data \cite{QUI}. Similarly we took $\Delta s=\Delta \bar{s}$ 
and we related them to ($\Delta \bar{d}-\Delta \bar{u}$), to agree with the second Bjorken sum rule.\\
The {\it eight} free parameters in Eqs.~(1,2) ($A$, $\bar{A}$ and $A_G$ are 
fixed by the normalization conditions $u-\bar{u}=2$, $d-\bar{d}=1$ and by 
the proton momentum sum rule) were
determined at the input scale from the comparison with a selected set of
very precise unpolarized and polarized DIS data from NMC, BCDMS, E665, ZEUS, H1
for $F^{p,d}_2(x,Q^2)$, from CCFR  for 
$xF^{\nu N}_3 (x,Q^2)$ and from SMC, E154 and E155 for
$g^{p,d,n}_1(x,Q^2)$.\\
We obtained
\begin{equation}
\bar{x}=0.09907,~ b=0.40962,~\tilde{b}=-0.25347~,
\label{equation5}
\end{equation}
\begin{equation}
X^+_{0u}=0.46128,~X^-_{0u}=0.29766,~X^-_{0d}=0.30174,~X^+_{0d}=0.22775~,
\label{equation6}
\end{equation}
\begin{equation}
\tilde{A}=0.08318,~A = 1.74938,~\bar{A} = 1.90801,~A_G = 14.27535~.
\label{equation7}
\end{equation}
In Eqs.~(1,2) the factors $X^{h}_{0q}$ and $(X^{-h}_{0q})^{-1}$ introduced in the numerators 
of the $q$'s and $\bar{q}$'s distributions, imply a modification of the
quantum statistical form, we were led to introduce to agree with 
experimental data. As we will see now, the justification of these factors
can be understood by studying the role 
of the transverse degree of freedhom for the partons.\\
Let us denote $p_i(x,p_T^2)$ \footnote{Here $i$ stands for the quark flavour and for
the spin component of the partons; $p_i(x,p_T^2)$ 
has now the dimension of GeV$^{-2}$.}
 a parton (quark or gluon) distribution with
transverse momentum $p_T$, which is assumed to be much smaller than its
longitudinal momentum $xp_z$, where $p_z$ is the momentum of the proton
with mass $M$.
We first consider the momentum sum rule, which reads now
\begin{equation}
\sum_i \int_0^1dx \int_0^{(p_T^2)_{max}} x p_i(x,p^2_T) dp^2_T = 1~,
\label{equation8}
\end{equation}
where the upper limit for the  $p^2_T$ integral, $(p_T^2)_{max}$, will be determined below. 
Next, we consider, the condition for the energy sum rule, analogous to Eq.~(8), where we assume that
the proton mass is much smaller than its momentum, an approximation which
holds in the deep inelastic regime.  Consequently, the proton energy is
$p_z + M^2/2p_z$, similarly the energy of a massless parton is $xp_z + p^2_T/2xp_z$, so this energy sum rule reads 
\begin{eqnarray} 
p_z\sum_i \int_0^1dx \int_0^{(p_T^2)_{max}} x p_i(x,p^2_T) dp^2_T &+& 
\nonumber\\
\sum_i \int_0^1dx \int_0^{(p_T^2)_{max}} 
p_i(x,p^2_T) \frac{p^2_T}{2xp_z}dp^2_T
&=&p_z+ \frac{M^2}{2p_z}~.
\label{equation9}
\end{eqnarray}
By using Eq.~(8) in Eq.~(9), it is easy to derive the constraint
\begin{equation}
\sum_i \int_0^1dx \int_0^{(p_T^2)_{max}} p_i(x,p^2_T)\frac{p^2_T}{x}dp^2_T = M^2 ~.
\label{equation10}
\end{equation}
By comparing Eq.~(10) and Eq.~(8), where it is clear that $x$ is bounded by 1, we see
that $p^2_T/x$ is bounded by $M^2$, so $(p_T^2)_{max}= xM^2$ .\\
The above sum rule will be used to determine the $p_T$ dependence, as we will explain now.
We recall the general method of statistical thermodynamics \cite{ES}, to find the most
probable occupation numbers $n_i$, for the energy levels $\epsilon_i$, when the total
energy of $N$ distinguishable particles is $E$. Since we have
\begin{equation}
N=\sum_i n_i ~~~\mbox{and}~~~~ E=\sum_i n_i \epsilon_i ~,
\label{equation11}
\end{equation}
one should look for the maximum of
\begin{equation}
\mbox{ln}[N!/\prod_i n_i!]+ \alpha (N-\sum_i n_i)+\beta (E-\sum_i n_i \epsilon_i)~,
\label{equation12}
\end{equation}
with the Lagrange multipliers method and we find
\begin{equation}
n_i= \exp(-\alpha - \beta \epsilon_i)~.
\label{equation13}
\end{equation}
The Lagrange multipliers $\alpha$ and $\beta$ are fixed by the two above
constraints Eq.~(11). Now by putting in correspondence $n_i$ with $p_i(x,p^2_T)$,
$N$ with 1 and $\alpha$ with $1/\bar x$, for the sum rule (8), and $E$ with $M^2$,
$\epsilon_i$ with $p^2_T/x$ and $\beta$ with $1/\mu^2$, for the sum rule (10), one
finds, in correspondence with Eq.~(13)
\begin{equation}
p_i(x,p^2_T)=\exp({\frac{-x}{\bar{x}}}+{\frac{-p^2_T}{x \mu^2}})~.
\label{equation14}
\end{equation}
$\mu$ is a parameter which has the dimension of a mass and it will be
determined later. We recall that
Eq.~(14) describes well the $p_T$ behavior of the particles 
produced in very high energy hadron-hadron scattering \cite{BP}.
For fermions, the Boltzmann exponential form Eq.~(14) is expected to be modified
into the product of two Fermi-Dirac expressions.\\
For the sake of clarity, we will first ignore  the $p_T$ dependence of the diffractive
contribution of quarks, antiquarks (second terms in Eqs.~(1,2)) and of the gluons, which
dominate the very low $x$ region. It certainly corresponds to very small $p_T$ values, since
$(p_T^2)_{max}= xM^2$ as seen above. So let us consider now 
the non-diffractive terms and instead of the helicity, as in Eqs.~(1,2), the spin
component of the partons along the momentum of the proton, which coincides with helicity
only when $p_T=0$.
The quantum statistics distributions for quarks and antiquarks read in this case
\begin{equation}
xq^{S_z}(x,p_T^{2})=\frac{F(x)}{[\exp[(x-X^{S_z}_{0q})/\bar{x}]+1]}
\frac{1}{[\exp[(p^2_T/x\mu^2-Y^{S_z}_{0q})/\bar x]+1]}~,
\label{equation15}
\end{equation}
\begin{equation}
x\bar{q}^{S_z}(x,p_T^{2})=\frac{{\bar F}(x)}{[\exp[(x+X^{-S_z}_{0q})/{\bar{x}}]+1]}
\frac{1}{[\exp[(p^2_T/x\mu^2+Y^{-S_z}_{0q})/\bar{x}]+1]},
\label{equation16}
\end{equation}
where $F(x)$ and ${\bar F}(x)$ do not depend on spin and flavour,
$X$ and $Y$ are the thermodynamical potentials related to the sum rules (8) and
(10), respectively \footnote{Here for convenience, we have divided by $\bar{x}$ 
the argument of the exponential
in the Fermi-Dirac expression of the $p_T$ dependence, in agreement with
the expression for the $x$ dependence.}. 
At high $p_T$, one has a Gaussian behavior, with a width
proportional to $\sqrt{x}$. It is important to note that this is at variance with
the usual fatorization assumption of the dependences in $x$ and $p_T$ \cite{AAM,SSY}.
We observe that the upper limit
of the integral in $x$, means that one parton is taking all the proton momentum.
Similarly $(p_T^2)_{max}=xM^2$ for the $p^2_T$ integral corresponds to the situation
where one parton is taking all the transverse energy. Therefore to 
simplify the formulas, we shall take as upper limit infinity, since the contribution
of the tail is negligible. In fact when one now integrates over $p^2_T$, the
right-hand side of Eq.~(15), one finds
\begin{equation}
\int_0^\infty \frac{dp^2_T}{\exp[(p^2_T/x\mu^2-Y^{S_z}_{0q})/\bar{x})]+1} = -
x \mu^2\bar{x} \mbox{Li}_1(-\exp[Y^{S_z}_{0q}/\bar{x}])~. 
\label{equation17}
\end{equation}
Here $\mbox{Li}_1$ denotes the polylogarithm function of order 1, which is known to arise
from the integral of Fermi-Dirac distributions and is such that
\begin{equation}
-\mbox{Li}_1(-\mbox{e}^y)= \int_0^\infty \frac{d\omega}{\mbox{e}^{(\omega-y)}+1}=
\ln{(1+\mbox{e}^{y})}~.
\label{equation18}
\end{equation} 
It is reasonable to assume the proportionality relationship
\begin{equation}
Y^{S_z}_{0q}=kX^{S_z}_{0q}~.
\label{equation19}
\end{equation}
It implies that the partons with a larger contribution to their first moments from the 
non-diffractive part, not only have a broader $x$ dependence, but also have 
a broader $p^2_T$ dependence, at every $x$.
We recall that $\mbox{Li}_1$ is an increasing function of its argument and we see from Eq.~(18) that
for large values of $y$, it becomes approximately proportional to $y$ and one has the high degeneracy of the
Fermi gas. Therefore this fully justifies, the phenomenological assumption 
of the proportionality to $X^{h}_{0q}$ made in Ref.~\cite{BBS1}, as seen in Eq.~(1).\\
By taking in Eq.~(15) \footnote{We identify $X_{0q}^{S_z}$ with $X_{0q}^{h}$ given in Eq.~(6)
and similarly for the potentials $Y^{S_z}_{0q}$.}
\begin{equation}
F(x) = -\frac{A x^{b-1} X^{+}_{0u}}{\mbox{Li}_1(-\exp[Y^{+}_{0u}/\bar{x}])\mu^2\bar{x}}~,
\label{equation20}
\end{equation}
we recover the first term in Eq.~(1) for the $u^{+}$ quark, which is the
dominating parton at large $x$. Eq.~(20) implies for the other quarks, that instead
of $X_{0q}^h$, one should use the factor $X_{0u}^{+}\mbox{Li}_1(-\exp[Y_{0q}^h/\bar{x}])/\mbox{Li}_1(-\exp[Y_{0u}^{+}/\bar{x}])$. \\
As long as the non-diffractive term for the antiquarks, by integrating
on $p^2_T$, one finds the factor $\mbox{Li}_1(-\exp[-Y^{h}_{0q}/\bar{x}])$.
At large negative $y$, $\mbox{Li}_1$ becomes proportional to $\exp{(-y)}$, which corresponds to the Boltzman
limit. Like for the quarks, we will fixe the normalization using the largest distribution,
namely $\bar{d}^-$, so similarly to Eq.~(20) we will take in Eq.~(16)
\begin{equation}
\bar{F}(x) = -\frac{\bar{A} x^{2b-1} (X^{+}_{0d})^{-1}}{\mbox{Li}_1(-\exp[-Y^{+}_{0d}/\bar{x}])\mu^2\bar{x}}~.
\label{equation21}
\end{equation}
We recover the first term in Eq.~(2) for $\bar{d}^-$ and for the other antiquarks instead of
$(X_{0q}^h)^{-1}$, one should use  $(X_{0d}^{+})^{-1}\mbox{Li}_1(-\exp[-Y_{0q}^h/\bar{x}])/\mbox{Li}_1(-\exp[-Y_{0d}^{+}/\bar{x}])$.
Now, in order to determine $k$, we require no change for the quark distributions, once integrated over $p_T^2$. This
can be achieved by choosing
\begin{equation}
k=1.42~.
\label{equation22}
\end{equation}
This value indicates that the potentials $Y_{0q}^h$ are of the same order as $X_{0q}^h$ and moreover
it implies lower values for the non-diffractive contributions of $\bar{u}^+$, $\bar{d}^+$ and $\bar{u}^-$.
Since the antiquark distributions are dominated by the diffractive contribution, these small modifications
are still consistent with the data.\\
At this stage we can try to use the energy sum rule Eq.~(10) to determine $\mu$, the
only unknown parameter, so far. We find to a very good approximation, the
remarkable value,
\begin{equation}
\mu = 1 \mbox{GeV}~.
\label{equation23}
\end{equation}
Moreover the energy sum rule is found to be saturated by a quark contribution
of 98\%, whereas the antiquark contribution is only 2\%. This value of $\mu$ was found by neglecting
the gluon and the diffractive contributions and it is clear that if they turn out to be important, $\mu$ will be smaller.\\
For completeness we now turn to the universal diffractive contribution to quarks and antiquarks in Eqs.~(1,2), namely
$xq^D(x,Q^2_0)=\tilde{A}x^{\tilde{b}}/[\exp(x/\bar{x}) + 1]$. Since $\tilde{b}<0$ (see Eq.~(5)), the introduction
of the $p_T$ dependence cannot be done similarly to the non diffractive contributions, because in the
energy sum rule Eq.~(10), it generates a singular behavior when $x \to 0$. Therefore we propose to
modify our prescription by taking at the input energy scale
\begin{equation}
xq^D(x,p^2_T)=\frac{\tilde{A}x^{\tilde{b}-2}}{\mbox{ln}(2)\mu^2\bar x}
\frac{1}{[\exp(x/\bar{x}) + 1]}\frac{1}{[\exp(p^2_T/\bar{x}x^2\mu^2)+1]}~,
\label{equation24}
\end{equation}
whose $p_T$ fall off is stronger, because $x\mu^2$ is now replaced by $x^2\mu^2$. Note that this is properly
normalized to recover $xq^D(x,Q^2_0)$ after integration over $p_T^2$. We have checked that $xq^D(x,p^2_T)$ gives
a negligible contribution to Eq.~(10), as expected, and we are led to a similar conclusion for the gluon, which has
the same small $x$ behavior. However since the gluon is parametrized by a quasi Bose-Einstein function, one
has to introduce a non-zero potential $Y_G$, to avoid the singular behavior of $\mbox{Li}_1(\exp[-Y_G/\bar{x}])$, when
$Y_G=0$ (see Eq.~(18)). The value of $Y_G$ is not constrained, but by taking a very small $Y_G$, it does not affect the energy sum rule.
Of course, one may ask why the gluon contributes so much to the momentum sum rule, Eq.~(8), and so little to the
energy sum rule, Eq.~(10). Using a better regularization procedure, we might find a larger contribution and consequently
 a smaller $\mu$ value. For this reason, we give the $p_T$ dependence as a function of the dimensionless variable
 $p_T^2/\mu^2$.\\
For illustration, we display in Fig.~1, $x{\mu^2}u(x,p_T^2/\mu^2,Q_0^2)$, the predicted dimensionless $u$-quark
parton distribution at input energy scale, versus $p_T^2/\mu^2$ for different
$x$ values. Note that the curves are limited to the value $(p_T^2)_{max}= xM^2$ and the Gaussian behavior in the high
$p_T$ region follows a rather flat behavior for $10^{-2}<p_T^2/\mu^2 < 10^{-1}$, or so. This feature is a natural
consequence of the statistical approach and it coincides, for a mean $x$ value, with the simplifying assumption
made in recent papers (see e.g. [13]) without real justification.
For $p_T^2/\mu^2 < 10^{-2}$, the very low $x$ region is strongly enhanced by the diffractive contribution, due
to the factor $x^{\tilde{b}-2}$ (see Eq.~(24)). However, since the regularization procedure proposed above is not unique, 
it should not be considered as a definite prediction.

\begin{figure}[htb]
\begin{center}
\leavevmode {\epsfysize=13.5cm \epsffile{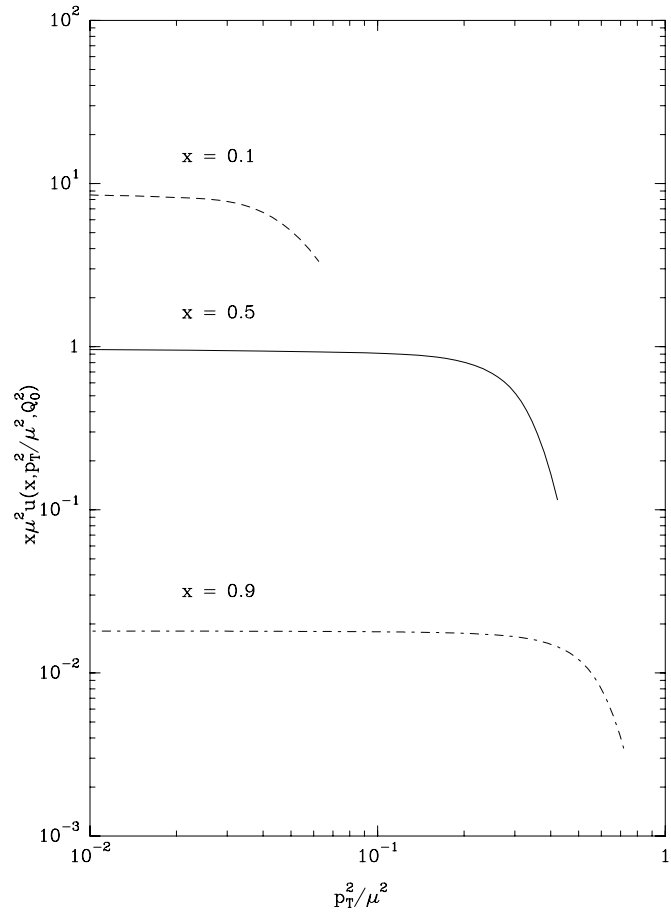}}
\end{center}
\vspace*{-14mm}
\caption[*]{\baselineskip 1pt
The predicted $x{\mu^2}u(x,p_T^2/\mu^2,Q_0^2)$ statistical distribution, versus $p_T^2/\mu^2$ for different values of $x$.}
\label{fi:pdfu}
\vspace*{-1.5ex}
\end{figure}

The theoretical fundation of the statistical parton distributions
proposed earlier in Ref.~\cite{BBS1} is strengthened by the present extension to the transverse degrees of
freedhom.  Indeed the energy sum rule Eq.~(10) cuts
naturally high $p^2_T$ values and this is very appealing, because an increasing phase-space for 
$p^2_T$ linear in $Q^2$ would support the absence of quantum statistical
effects for diluted fermions and bosons. Our predicted light 
partons distributions $p_i(x,p^2_T)$, both unpolarized and polarized given by Eqs.~(15,16), remain
to be checked against appropriate experimental data, sensitive to some effects arising from the transverse momentum
of partons inside the proton. Here it is important to emphasize
that the $p_T$ dependence is only valid in a limited kinematic region, say
$0.1<x<0.9$, which is not dominated by the diffractive contribution. Clearly
in this large $x$ region, we predict the mean value of $p_T^2$, namely $<p_T^2>$, to
increase with $x$, a behaviour which is at variance with what one expects
in the very small $x$ region from the BFKL evolution, assuming the emission of
a hudge number of gluons \cite{lowx}. However, so far there is no experimental
evidence for the BFKL evolution.

 Finally, we note that our procedure restricts to rotational invariance, the
transverse momentum dependence of the parton distributions. The next step should be to introduce some
azimuthal dependence, which is needed to study the important topic of single-spin asymmetries \cite {BMP}.

\end{document}